\documentclass{article}
\usepackage{amsmath}
\usepackage{fancyhdr}

\input{tcilatex}

\begin{document}

\title{An Improved Quantum Scheduling Algorithm}
\author{Lov K. Grover \thanks{%
This research was partly supported by NSA\ \&\ ARO under contract no.
DAAG55-98-C-0040.} \and \textit{lkgrover@bell-labs.com} \\
1D435 Bell Laboratories, Lucent Technologies,\\
600-700 Mountain Avenue, Murray Hill, NJ 07974}
\date{}
\maketitle

\begin{abstract}
The scheduling problem consists of finding a common 1 in two remotely
located $N$\ bit strings. Denote the number of 1s in the string with the
fewer 1s by $\epsilon N$. Classically, it needs at least $O(\epsilon N\log
_{2}N)$ bits of communication to find the common 1. The best known quantum
algorithm would require $O(\sqrt{N}\log _{2}N)$ qubits of
communication.\thinspace This paper gives a quantum algorithm to find the
common 1 with only $O(\sqrt{\epsilon N}\log _{2}N)$ qubits of communication.
\end{abstract}

\baselineskip=14pt

\newpage \pagebreak

\section{Introduction}

Alice \&\ Bob each have a calendar of their appointments for $N$ slots of
time. They need to find a common slot for a meeting when each of them is
available. How many bits of information do they need to exchange? This
problem is also known as the intersection problem, i.e. the problem is to
find a common $1$ in two remotely located strings.

It has been shown that if Alice \&\ Bob were exchanging classical bits, they
would need to exchange $O(N)\;$bits of information \cite{yao} (the intuition
is that if they were to try to find the answer while exchanging fewer bits,
they might leave out information about the common slot).

It was a surprising result when Buhrman, Cleve \&\ Wigderson (BCW) \cite{bcw}
discovered a quantum mechanical technique through which Alice \&\ Bob could
identify the common slot while exchanging only $O(\sqrt{N}\log N)$ qubits.
This was surprising because there is a well known result called Holevo's
Theorem which proves that a single qubit cannot carry more than one bit of
classical information \cite{holevo}. It takes some thought to realize that
the BCW result does not violate Holevo's Theorem since Alice \&\ Bob could
not use it to transmit $N$ bits of information. What the BCW\ algorithm
accomplishes is to reduce the communication complexity of the scheduling
problem from $O(N)$ classical bits to only $O(\sqrt{N}\log N)$ quantum
mechanical qubits. This was the first significant reduction in communication
complexity achieved through quantum communication.

The basis of the algorithm of \ BCW was to carry out a quantum search on the 
$N$ possible slots for the common slot. Since Alice \&\ Bob each have part
of the information, this requires a distributed search with the qubits being
transferred back and forth as described in section 3. This requires $O(\sqrt{%
N}\log N)$ qubits of communication which is considerably fewer than the $%
O(N) $ qubits of classical communication that would be required.

As described above, the number of qubits of communication required is
approximately the square-root of the number of bits of classical
communication required in the most general case. However, in the special
case when the string of either Alice or Bob has few 1s, it is possible to
design a classical algorithm requiring much less communication. For example,
if Alice's string has $\epsilon N$ 1s, she could encode the position of
these using $\epsilon N\log _{2}N$ bits and send these to Bob who could then
find the common slot. This paper gives a modification to the BCW algorithm
so that it gives a square-root advantage over the best classical algorithm
even when the strings of Alice or Bob have few 1s.

\section{The Quantum Search Algorithm}

In order to describe the algorithm of BCW, we first need to describe the
quantum search algorithm \cite{grover96}. There are only three operations
required by the search algorithm:\ $W,\;I_{\overline{0}},I_{t}.$These are
described below.

$I_{t}$ is a selective inversion of the target state. This can be achieved
provided we have a quantum mechanical black box that can evaluate whether or
not a given state is the target state - note that this does not need any a
priori knowledge of which the target state is. See \ref{brassard} for a
quantum circuit that accomplishes \ a selective phase inversion of $t$ using
such a black box.

$I_{\overline{0}}$ is the selective inversion of the $\overline{0}$ state
(i.e. the state in which all qubits are $0$). $W$ is the Walsh-Hadamard
Transformation.

\bigskip The quantum search algorithm showed that, if we start from $%
\overline{0}$ and carry out $\frac{\pi \sqrt{N}}{4}$ repetitions of the
sequence of operations $I_{\overline{0}}WI_{t}W,$ followed by $W$, we reach
the $t$ state with certainty. Equivalently:%
\begin{equation*}
W\underset{\frac{\pi \sqrt{N}}{4}repetitions}{\underbrace{\left( I_{%
\overline{0}}WI_{t}W\right) \ldots \left( I_{\overline{0}}WI_{t}W\right)
\left( I_{\overline{0}}WI_{t}W\right) \left( I_{\overline{0}}WI_{t}W\right) }%
}\left| \overline{0}\right\rangle =\left| t\right\rangle
\end{equation*}

A\ measurement after this will reveal which the $t$ state is. Note that this
requires only $O(\sqrt{N})$ operations.

\section{Distributed Searching}

In the scheduling problem, part of the information is with Alice \&\ part of
it with Bob. The insight of BCW was to observe that the quantum search
algorithm could still be carried out by transferring the qubits back \&\
forth. There are only three operations required by the search algorithm:\ $%
W,\;I_{\overline{0}},I_{t}.$ The first two are independent of the $t$ state
and can be carried out anywhere ($I_{\overline{0}}$ requires all qubits to
be available together).

$I_{t}$ clearly depends on the solution the information about which is with
Alice \&\ Bob. It requires that both Alice's and Bob's schedules be
satisfied. A modification of the circuit shown in the figure in the previous
section accomplishes this. It requires part of the function to be evaluated
by Alice \&$\;$the resulting qubits to be passed to Bob who evaluates the
rest of the function. This ensures that both Alice's \&\ Bob's functions
evaluate to 1. Please see \cite{bcw} for details.

\section{\protect\bigskip\ Amplitude Amplification}

A few years after the invention of the quantum search algorithm, it was
generalized to a much larger class of applications known as the amplitude
amplification algorithms \cite{ampt. amp.} (similar results are
independently proved in \cite{bht}). In these algorithms, the amplitude
produced in a particular state by a unitary operation $U$, can be \textit{%
amplified} by successively repeating the sequence of operations:\ $%
Q=I_{s}U^{\dag }I_{t}U$. It was proved that if we start from the $s$ state
and repeat the operation sequence $I_{s}U^{\dag }I_{t}U,$ $\eta $ times
followed by a single repetition of $U$, then the amplitude in the $t$ state
becomes approximately $\eta U_{ts}$ (provided $\eta U_{ts}\ll 1$). Also, if
we start from $s$ and carry out $\frac{\pi }{4\left| U_{ts}\right| }$
repetitions of $Q$ followed by a single repetition of $U,$ we reach $t$ with
certainty.

The quantum search algorithm was a particular case of amplitude
amplification with the Walsh-Hadamard Transformation being the $U$ operation
and $s$ being the $\overline{0}$ state. For any $t$, $\left| U_{ts}\right| =%
\frac{1}{\sqrt{N}}.$ It follows from the amplitude amplification principle
that if we start from $\overline{0}$ and carry out $\frac{\pi \sqrt{N}}{4}$
repetitions of the sequence of operations $I_{\overline{0}}WI_{t}W,$
followed by $W$, we reach the $t$ state with certainty.

In this paper we use the amplitude amplification principle for developing a
new scheduling algorithm. This is achieved by designing a sequence of
transformations that produce an amplitude of $\frac{1}{\sqrt{\epsilon N}}$
in the $t$ state and much smaller amplitudes in other states while requiring 
$\log _{2}N$ qubits of communication. Therefore by the amplitude
amplification principle, in $\frac{\pi }{4}\sqrt{\epsilon N}$ repetitions of
this transformation, we can concentrate most of the amplitude in the $t$
state.

\section{Improved Scheduling Algorithm}

The following algorithm assumes a single common 1 in the two strings. The
algorithm is easily extended to the situation with multiple common 1s using
standard approaches, e.g. \cite{brassard}. Alice \&\ Bob each count the
number of 1's in their strings (or equivalently the number of slots they are
each available for). They exchange this information using $\log _{2}N$ bits
of communication.\ The person with the fewer 1s (say Alice)\ starts.

Assume that Alice has $\epsilon N$ 1's in her $N$ bit string. She starts
with a register of $\log _{2}N$ qubits that encode the $N$ slots. She starts
with these in the $\overline{0}$ state. Next consider the transformation
consisting of $\frac{\pi }{4}\sqrt{\epsilon N}$ applications of the quantum
search operator $I_{\overline{0}}WI_{A}W,$ followed by $W$ (here $I_{A}$
inverts the amplitude of each of the states when Alice is available). This
produces a superposition concentrated in states in which she is available.
Denote this composite transformation that transforms $\overline{0}$ into the
superposition corresponding to her available slots, by $U$, i.e.%
\begin{equation*}
U\equiv W\underset{\frac{\pi \sqrt{\epsilon N}}{4}repetitions}{\underbrace{%
\left( I_{\overline{0}}WI_{A}W\right) \ldots \left( I_{\overline{0}%
}WI_{A}W\right) \left( I_{\overline{0}}WI_{A}W\right) \left( I_{\overline{0}%
}WI_{A}W\right) }}
\end{equation*}%
Note that $U^{\dag }$ consists of the application of the adjoints of the
operations that constitute $U$ but in the opposite order, i.e.

\begin{equation*}
U^{^{\dag }}\equiv W\underset{\frac{\pi \sqrt{\epsilon N}}{4}repetitions}{%
\underbrace{\left( I_{A}WI_{\overline{0}}W\right) \ldots \left( I_{A}WI_{%
\overline{0}}W\right) \left( I_{A}WI_{\overline{0}}W\right) \left( I_{A}WI_{%
\overline{0}}W\right) }}
\end{equation*}

Next apply the following sequence of transformations:%
\begin{equation*}
U\;\underset{\frac{\pi \sqrt{\epsilon N}}{4}repetitions}{\underbrace{\left(
I_{\overline{0}}U^{^{\dag }}I_{B}U\right) \ldots \left( I_{\overline{0}%
}U^{^{\dag }}I_{B}U\right) \;\left( I_{\overline{0}}U^{^{\dag
}}I_{B}U\right) }}\left| \overline{0}\right\rangle
\end{equation*}%
Note that this needs only $\frac{\pi \sqrt{\epsilon N}}{2}\log _{2}N$ qubits
of communication since Alice can carry out all but the $I_{B}$ operations
for which the register needs to be sent to Bob and returned. The number of
times the register needs to be sent to Bob is equal to the number of $I_{B}$
operations.

It follows by the amplitude amplification principle that if we start from $%
\left| \overline{0}\right\rangle ,$then after $\frac{\pi }{4\sum_{B}\left|
U_{B\overline{0}}\right| ^{2}}$ repetitions of the $I_{B}UI_{\overline{0}%
}U^{^{\dag }}$ followed by a single application of $U$, all the amplitude is
concentrated in the states inverted by the $I_{B}$ operation (i.e. the set
of states that satisfy Bob's schedule) with amplitude in each $B$ state
proportional to $U_{B\overline{0}}$. The amplitude $U_{B\overline{0}}$, is 0
except in the states that satisfy Alice's schedule. Therefore the
superposition is entirely concentrated in states that satisfy both Alice's
\&\ Bob's schedule. For convenience we have assumed a single common slot,
the method of \cite{brassard} easily generalizes it to multiple common slots.

\section{Other Tradeoffs \&\ Applications}

The algorithm of this paper requires $O(\sqrt{\epsilon N})$ cycles, each
cycle requires $O\left( \sqrt{\frac{N}{\epsilon N}}\right) $ steps of
computation (assuming the busier person has $\epsilon N$ openings and each
selective inversion takes $O(1)$ steps) - which gives $O(\sqrt{N})$ total
steps of computation while requiring only $O\left( \sqrt{\epsilon N}\right) $
communication (ignoring log factors). This is in comparison to the BCW \cite%
{bcw} algorithm that requires $O\left( \sqrt{N}\right) $steps of computation
and $O\left( \sqrt{N}\right) $communication. As described above, this gives
our algorithm an advantage when the communication cost is significant. It is
interesting to speculate about the situation when Alice's \&\ Bob's
selective inversions take different numbers of steps - it might be possible
to design algorithms that are more efficient even in terms of computation,
or in terms of some combination of computation and communication.

Alternatively, one can extend this method to solve problems where the
problem is easiest posed as an intersection of two different queries, one of
which is more selective than the other.

\end{document}